\documentclass[fleqn,usenatbib]{mnras}

\usepackage[T1]{fontenc}

\DeclareRobustCommand{\VAN}[3]{#2}
\let\VANthebibliography\thebibliography
\def\thebibliography{\DeclareRobustCommand{\VAN}[3]{##3}\VANthebibliography}

\usepackage{graphicx}	
\usepackage{amsmath}	
\usepackage{amssymb}	
\usepackage{multirow}
\usepackage{booktabs, tabularx}
\usepackage{listings}

\usepackage{algorithmicx}
\usepackage{algorithm}
\usepackage{algpseudocode}

\usepackage{newtxtext,newtxmath}

\title[Classes of pulsars and the MST]{
Quantitative determination of minimum spanning tree structures:\\
Using the pulsar tree for analyzing the appearance of new classes of pulsars}

\author[García \& Torres]{
C. R. Garc\'{i}a$^{1,3}$\thanks{E-mail: crodriguez@ice.csic.es}, 
Diego F. Torres$^{1,2,3}$\thanks{E-mail: dtorres@ice.csic.es}
\\
$^{1}$Institute of Space Sciences (ICE, CSIC), Campus UAB, Carrer de Can Magrans s/n, 08193 Barcelona, Spain\\
$^{2}$Institució Catalana de Recerca i Estudis Avançats (ICREA), E-08010 Barcelona, Spain \\
$^{3}$Institut d’Estudis Espacials de Catalunya (IEEC), 08034 Barcelona, Spain
}

\pubyear{2023}

\begin{document}
\label{firstpage}
\pagerange{\pageref{firstpage}--\pageref{lastpage}}
\maketitle

\begin{abstract}
In this work, we introduce a quantitative methodology to define what is the main trunk and what are the significant branches of a minimum spanning tree (MST). We apply it to the pulsar tree, i.e. the MST 
of the pulsar population constructed upon a Euclidean distance over the pulsar's intrinsic properties. Our method makes use of the betweenness centrality estimator, as well as of non-parametric tests to establish the distinct character of the defined branches. Armed with these concepts, we study how the pulsar population has evolved throughout history, and analyze how to judge whether a new class of pulsars appears in new data, future surveys, or new incarnations of pulsar catalogs.

\end{abstract}


\section{Introduction}
\label{intro}

In a recent work \citep{MST-1},  we have applied a principal components analysis, e.g., \cite{Pearson1901, Shlens2014}, over magnitudes depending on the intrinsic pulsar’s timing properties (considered as proxies to relevant physical pulsar features, e.g., the magnetic surface field, the spin-down power, etc.), to analyze whether the information contained by the pulsar’s period and period derivative are enough to describe the variety of the pulsar population. We showed that $P\dot{P}$ are not principal components and do not contain the full variance of the pulsar population. Thus, any distance ranking or visualization based only on $P$ and $\dot{P}$ is potentially misleading. Subsequently, we have introduced the use of graph theory to the problem, and in particular, presented 
the Pulsar Tree. This is the minimum spanning tree (MST, see e.g., \cite{Gower1969,Kruskal1956}) of the pulsar population constructed upon a Euclidean distance over the pulsar's intrinsic properties. We prepared as well an online tool {\url{http://www.pulsartree.ice.csic.es}}, a site which contains visualization tools and data to allow users to gather information in terms of the MST and the distance ranking. Here, we shall build upon \cite{MST-1} and we shall take for granted that the reader is aware of its introductory appendices accounting for the needed conceptual ingredients from graph theory. This work has the following aims:
\begin{itemize}
\item We want to establish a quantitative methodology to define what is the main trunk and what are the significant branches of a pulsar tree (or any MST, in general). In particular, we want to introduce a qualifier to consider whether the branches of the tree have statistically significant differences among themselves.

\item Once we establish what is the trunk and the significant branches, we want to develop a methodology to tell us whether these structures have grown 
in earlier incarnations of the pulsar population.

\item As a spinoff of the latter study, we analyze how to judge whether a new class of pulsars appears in new data, future surveys, or new incarnations of pulsar catalogs.

\end{itemize}

We shall use v1.67 of the ATNF catalog \citep{ATNF-Catalog},  collecting pulsars entered in the database until March 2022. This version contains 2509 pulsars, of which 2242 are isolated pulsars and 267 are pulsars belonging to binary systems having period derivatives greater than zero. Using the label 'date' of the ATNF catalog, which refers to the date of the pulsar discovery, we shall also consider the historical evolution of the pulsar population.

\section{Identifying the main trunk and branches of the tree}
\label{method}

\subsection{Betweenness centrality}
\label{idea}

Betweenness centrality defines how central a node is, or put otherwise, how many times a given node of a graph is in between any two others (see \cite{original}, see also \cite{Moxley1974}). More precisely, it measures centrality from the ratio between the number of times a node $v$ appears on the shortest path, also known as geodesic distance, between any two other nodes $(s, t)$, and the number of possible shortest paths that could occur between them (e.g., see \cite{Brandes}), 

\begin{eqnarray}
C_{B} = \frac{2}{(N-1)(N-2)}\sum_{s\neq v\neq t \in V} \frac{\sigma_{s,t}(v)}{\sigma_{s,t}}
~.\label{eq:betweenness_centrality}
\end{eqnarray}

\begin{figure*}
\centering
\includegraphics[width=\textwidth]{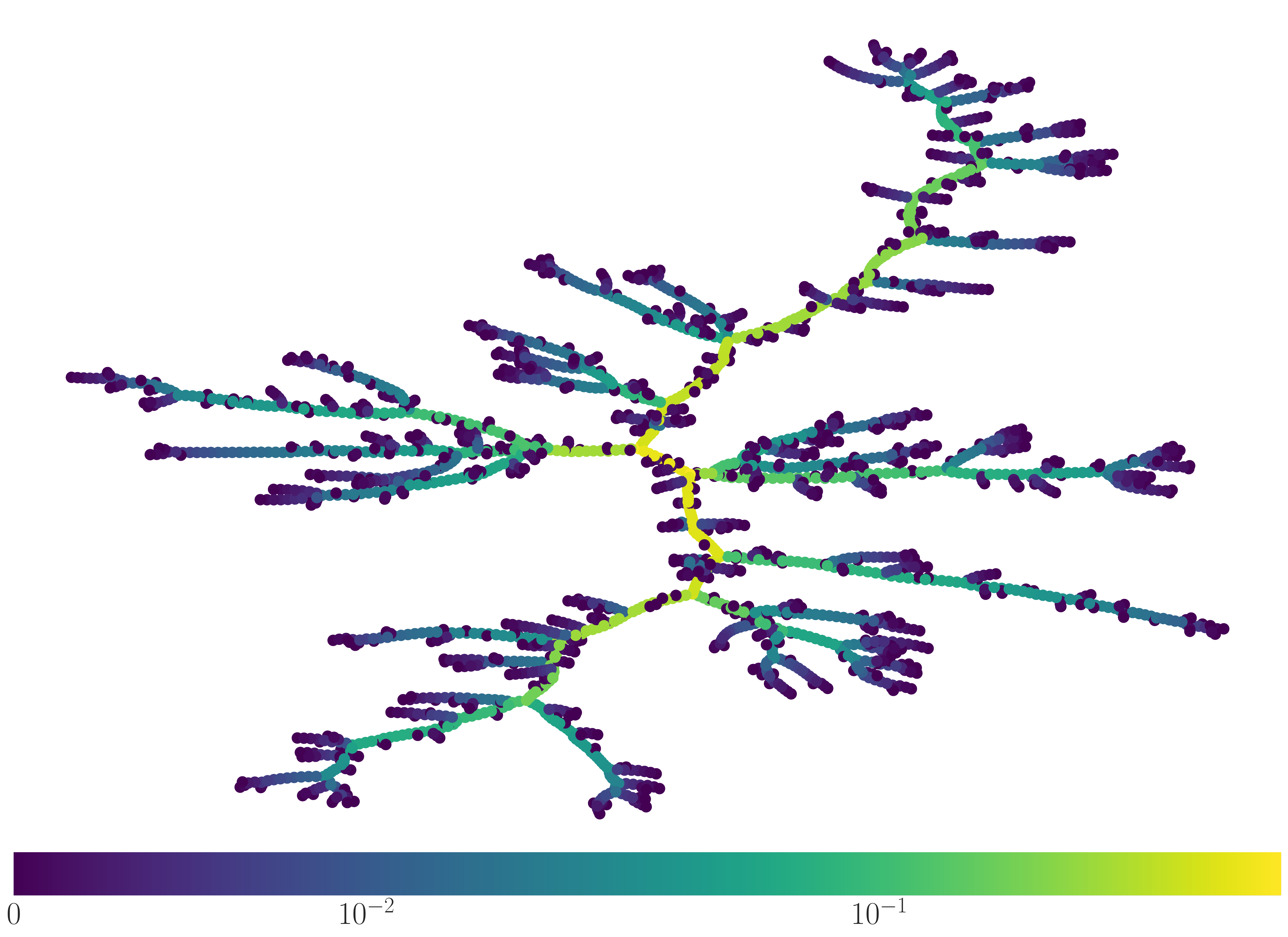}
\caption{A color-coded representation of MST(2509, 2508) according to the values of the betweenness centrality coefficient after application of Eq. (\ref{eq:betweenness_centrality}). 
The color intensity varies from zero (with the strongest color), for those nodes located at the termination of the graph being nodes of 1 degree, to the nodes located in the central part of the MST where $C_B$ reaches its highest values around 0.6. 
}
       \label{fig:BetCen_ModelVersion}
\end{figure*}

\begin{figure}
\centering
  \includegraphics[width=0.35\textwidth]{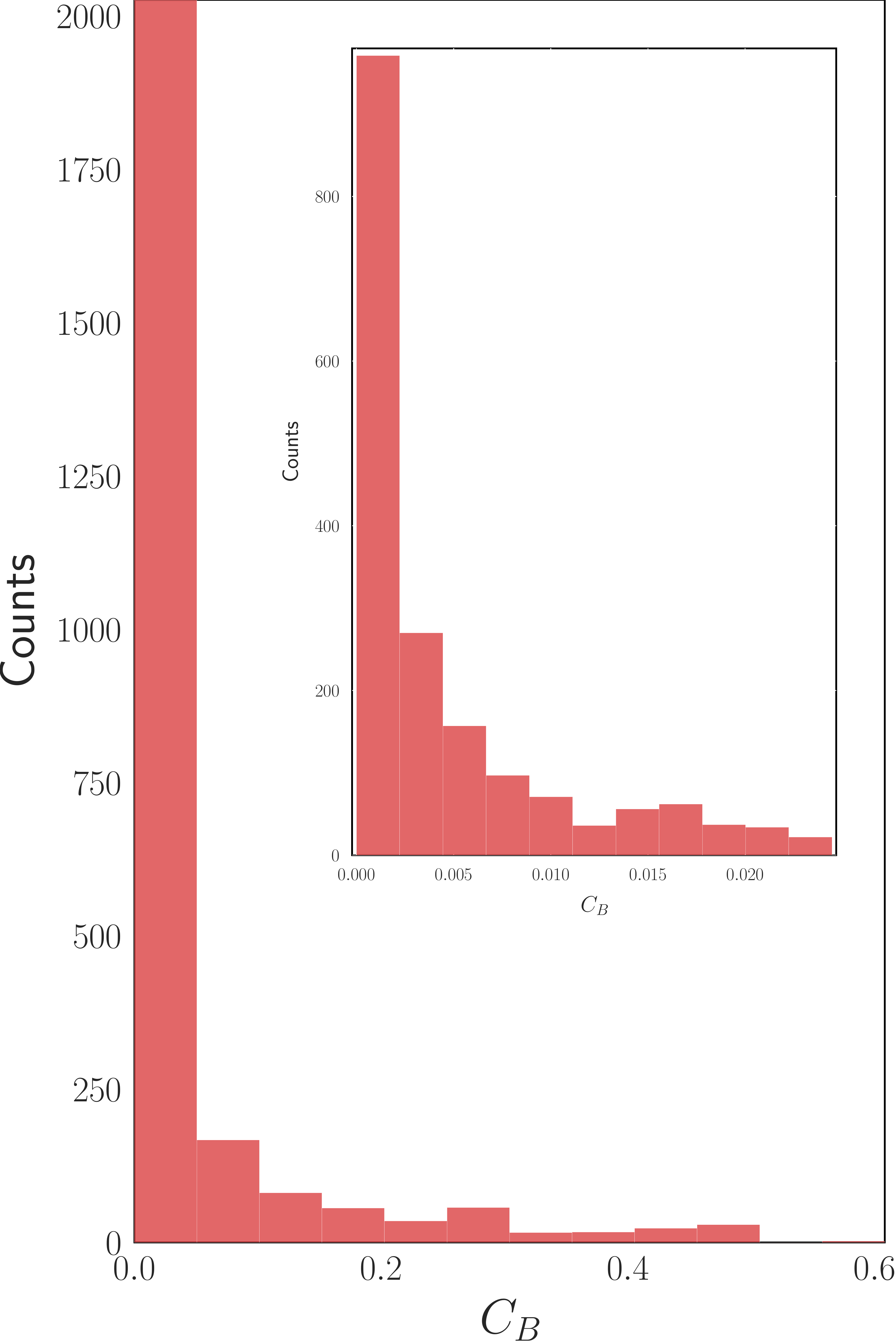}
  \caption{Distribution of the betweenness centrality  values seen after applying Eq. (\ref{eq:betweenness_centrality}) on the MST of the pulsar population (ATNF v.1.67,  shown in Fig. \ref{fig:BetCen_ModelVersion}). In the upper right corner, the distribution of about 70\% of the data is zoomed in. 
}
  \label{Distributions_betweenness}
\end{figure}

Here, $\sigma_{s,t}(v)$ can be equal to 1 or to 0. 
$\sigma_{s,t}(v) =1$ only when any geodesic distance between the nodes 
${s,t}$ pass through the node $v$. In a general graph, $\sigma_{s,t}$ is the total number of shortest paths between two nodes $(s,t)$. As imposed by the inexistence of closed loops in an MST, i.e., the MST is acyclic  (see eg., \cite{Tarjan1983}), only one path is possible between any two nodes of an MST (see eg., \cite{Wilson2010}). Thus,  the denominator within the sum of Eq. (1) equals unity, $\sigma_{s,t} =1$, for any two pairs of nodes, since there is only one path connecting two nodes in the MST. To avoid duplicities, and resorting to the non-directionality of the MST, $\sigma_{s,t} = \sigma_{t,s}$, and will be taken into account only once in the sum Eq. (\ref{eq:betweenness_centrality}) is normalized by multiplying it by $2/(N-1)(N-2)$ where $N=|V|$ is the number of nodes we have in the set $V$. This is done to compare results obtained between graphs of different sizes regarding the number of nodes. An explicit example with an MST, and the graph from which we obtain that MST, of a few nodes, will help clarify the computation, and we provide such an example in the Appendix. Betweenness centrality is in fact helpful in formalizing an obvious mental idea of how central a node is for a given graph, allowing us to have mathematical definitions based on its distribution. Fig. \ref{fig:BetCen_ModelVersion} shows the MST of the current pulsar population as presented in our previous work \citep{MST-1} after the application of Eq. (\ref{eq:betweenness_centrality}).
Fig. \ref{Distributions_betweenness} shows the distribution of the betweenness centrality values just computed. Outliers are clearly appearing in this asymmetric distribution.

\subsection{Definition of the main trunk }
\label{conditions}

To establish a central interval for the distribution of betweenness centrality we shall use an appropriate technique for asymmetric distributions, based on quartiles. The central interval will be defined as (see, e.g., \cite{Tukey}),

\begin{eqnarray}
[Q_{1}-k\times IQR];[Q_{3}+k\times IQR]
~,\label{eq:outliers}
\end{eqnarray}

where $k$ is a coefficient typically taken equal to 3. The use of quartiles (${Q_{1}}$, ${Q_{3}}$) to identify central values and outliers are validated for both symmetric and asymmetric distributions: they do not assume anything about the mean or the standard deviation, and their use is compatible with distributions of positive rightward skew, as ours. Furthermore, any asymmetric distribution is more robustly defined by the median as a measure of central tendency and the interquartile range (IQR= $Q_{3}-Q_{1}$) as a measure of its dispersion, as both are less sensitive to extreme values. For our case, the right-hand side of the central interval shown in Eq. (\ref{eq:outliers}), together with the usual value of 3 for $k$ sets the condition for a node to be considered central (i.e., an outlier of the betweenness centrality distribution). This condition is to be taken as necessary, but not sufficient for a node to be considered part of the main trunk. In addition to being formed by outliers of the betweenness centrality distribution (something that relates to a bare-eye identification as the main trunk in Fig. \ref{fig:BetCen_ModelVersion}) we need to provide a criterion to establish where it starts and ends. To that aim, we shall request a topological condition: As the trunk is a path, i.e., a sequence of consecutive nodes containing no duplicates, we shall require that it must be starting and ending at nodes whose degree must be greater than 2. In this way, we ensure that at the terminations of the trunk, there are nodes that give rise to {\it significant} substructure, i.e., to relate to the mind-image of branches opening up from the trunk (branches are to be defined more precisely below) with a possible physical significance. The term {\it significant} used above --which quantitative meaning is discussed next -- is herein added to explicitly avoid ending the tree in a degree 3 node but from where the branches departing from it are formed just by a few nodes (noise). Thus, we define the main trunk as the longest substructure of the MST formed by outliers of the betweenness centrality distribution, starting and ending in nodes of degree 3 or higher, and giving rise to significant branches.

\subsection{Significant branches}
\label{conditions-branch}

We adopt a general conceptual definition (that is useful for our pulsar tree case, as well as for any other MST): a relevant branch is defined as a group of nodes departing from the main trunk that contains at least a minimum percentage of the total population of nodes in the graph and can be quantitatively distinguished from other branches. To impose that a branch contains at least a minimum percentage of the total number of nodes in the graph (to be referred to as significance threshold) is done to avoid having branches with just a small collection of nodes, usually departing briefly from other substructures. These are to be considered noise of the latter. The significance threshold will be empirically found directly from the MST being studied. Note that the larger this threshold is (i.e., the more nodes will be assigned to branches), the smaller the main trunk will be. For instance, if this threshold is 10\%, i.e., each branch should contain at least 10\% of the total number of nodes, there will be fewer branches (and with more nodes each), and a smaller main trunk, than when it is fixed at 5\%. To define the significance threshold and measure at once whether one branch distinguishes from others we shall use the Kolmogorov-Smirnov (KS) statistics. The KS test compares two distributions under study via the distance between their empirical cumulative distribution functions, e.g., see \cite{KS-test4, KS-test5, KS-test3}. The KS test does not assume any form of distribution beforehand, which makes it a non-parametric test something that allows its use for any type of distribution. The aim here will be to see if we can reject the null hypothesis ($H_0$): the distribution of the properties of the nodes of two given branches are consistent with them coming from the same parent distribution. We will be seeking to reject this null hypothesis at 95\% confidence level (CL) or better. When this happens, we shall be establishing that whatever distance is used to compute the weights between nodes and form the MST is separating branches whose nodes are drawn from statistically distinct parent populations. Specifically for the case of the pulsar tree, we shall recall the results of our PCA analysis  \citep{MST-1}. It defined that two  principal components were needed for the set of variables studied  (spin period, spin period derivative, surface magnetic flux density at the equator, the magnetic field at the light cylinder, spin-down energy loss rate, characteristic age, surface electric voltage, and Goldreich-Julian charge density). The distributions for the principal components of the pulsar population $PC_1$ and $PC_2$ are skewed, asymmetric distributions with pronounced rightward and leftward shifts. The explained variance by the first principal component, $PC_1$, reaches $\sim 72$\% in the current, most numerous incarnation of the catalog. Thus, we shall request that the KS test rejects that any two branches have a $PC_1$ distribution that could come from the same parent population. The threshold value from which branches are defined can then be iteratively fixed to the minimum value possible at which all branches are distinct under the KS test applied to their $PC_1$ distribution. If the threshold would be smaller, the number of branches increases, the average branch size decreases, and not all the  smaller sub-structures represent distinct populations.

\subsubsection{Other tests and principal components}

One can entertain that several variations on the methodology above are possible in the general case. For instance, one can ask why not requesting that both $PC_1$ and $PC_2$ distributions are distinguished for all branches via the KS test. One can also ask why not using a different test from the KS. We briefly comment on these aspects below and justify the proposed approach as the most reasonable. When a principal component represents low values of explained variance, it means that the sample presents incomplete partial information in this dimension so all associated distributions will be of lower relevance for representing the nodes. Nevertheless, we find that our results are stable if we were to use both $PC_1$ and $PC_2$ to define the minimum threshold. We have checked that under $PC_2$, our branches distinguish themselves in all cases and that the KS test with $PC_2$ rejects the null hypothesis at smaller significant thresholds than  $PC_1$. This justifies our choice of using the most relevant principal component only. Regarding the second question posed, we find the KS test especially suited to the kind of null hypothesis we want to reject and simple enough for our aims. Other usual tests in astronomy, such as the F-test, e.g., see \cite{Ftests}, are not really applicable as if a distribution is skewed, one could argue that the variance is not an optimum dispersion measure. Other non-parametric tests for comparing two populations, such as the Anderson-Darling (e.g., see \cite{anderson}) are also based on empirical distribution functions. We have checked that in all cases where the KS test rejects the null hypothesis, also the Anderson-Darling test does. We prefer the KS test over the Anderson-Darling -- apart from its simplicity -- also because the KS test loses rejection power the smaller the samples being compared are. This is reasonable for our case since it does not make sense to define branches formed by just a few pulsars.

\subsection{The pulsar tree main trunk and branches}
\label{current branches}

\begin{figure*}
\centering
\includegraphics[width=\textwidth]{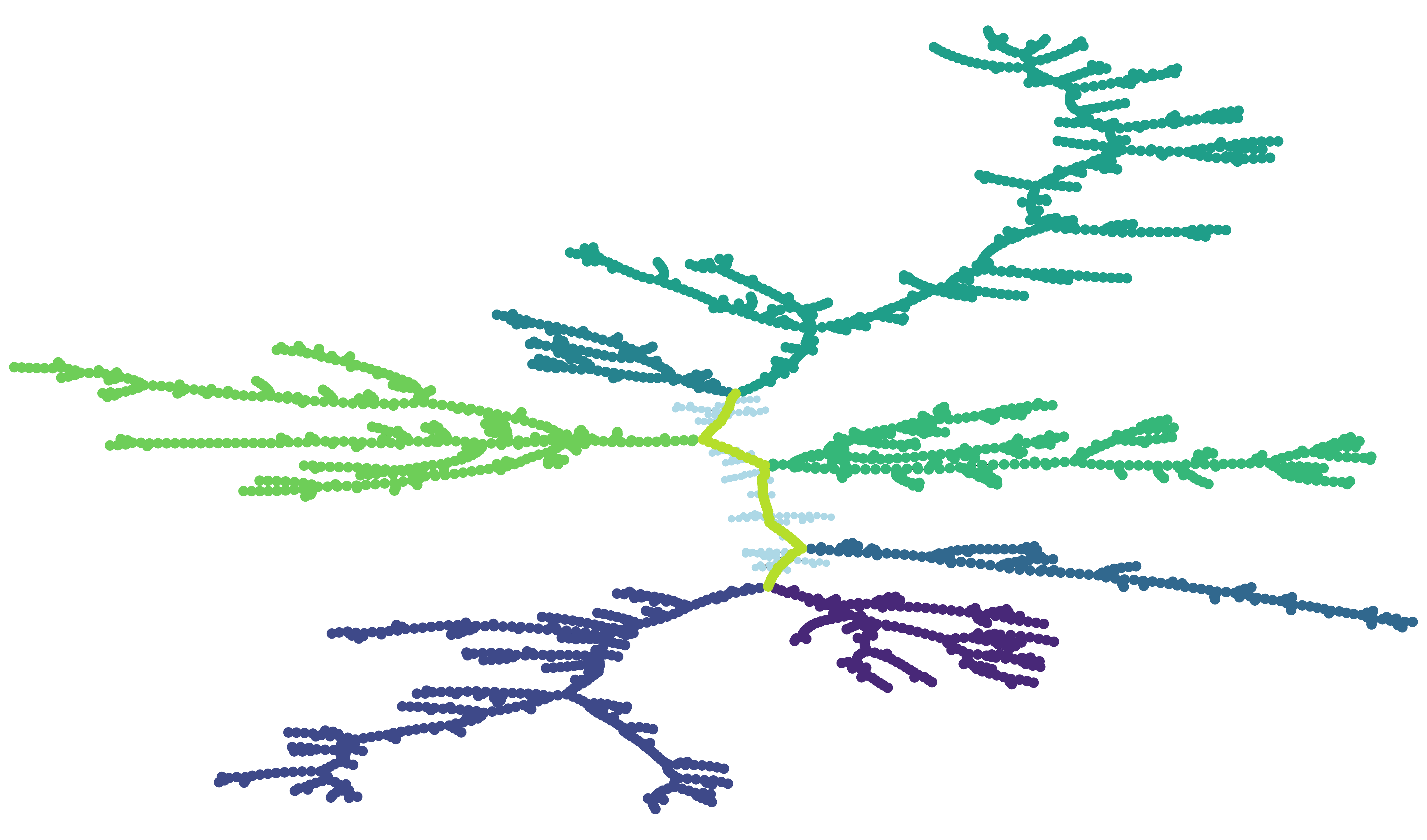}
\caption{Significant branches of v.1.67 (March 2022) are shown after applying the techniques described in the text. These have at least 3.8\% of the total population size, following the branches clockwise (from 12:00) they have the following number of pulsars:  585, 364, 157, 224, 444, 427, and 119, respectively. The MST nodes depicted in the main part are 64 and represent the trunk. Note that the remaining light blue (125) nodes belong to non-significant branches and are considered to be the noise of the main trunk. 
}
       \label{fig:MST_Branches_V167}
\end{figure*}

\begin{figure*}
  \centering
  \includegraphics[width=1\textwidth]{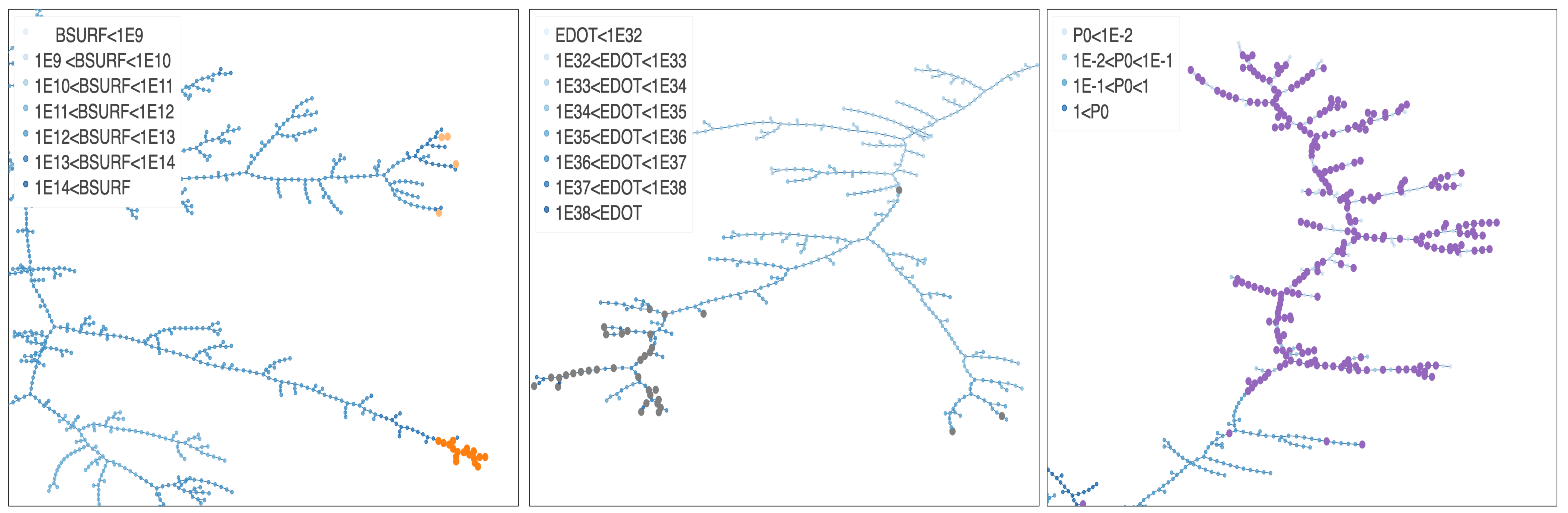}
  \caption{
Three examples of the distribution of variables for some of the significant branches of the pulsar tree, as can be extracted from the online tool at {\url{http://www.pulsartree.ice.csic.es}}. We suggest looking at these examples directly there  to have access to additional functionalities, being able to zoom in and see individual values for each pulsar. The left panel shows the distribution of the surface magnetic field in the middle branches, increasing towards the outskirts of the MST, ending in low-field magnetars (light orange) and classical magnetars (orange). The middle panel shows the spin-down power distribution of the bottom branch, with those having a pulsar wind nebula detected in TeV \citep{HESSPWNe} (grey) noted. The right panel shows the distribution of the period in the upper branches of the MST where binaries (purple) and long-period pulsars are located. 
  }
  \label{fig:pt-examples}
\end{figure*}

\begin{figure*}
  \includegraphics[width=1\textwidth]{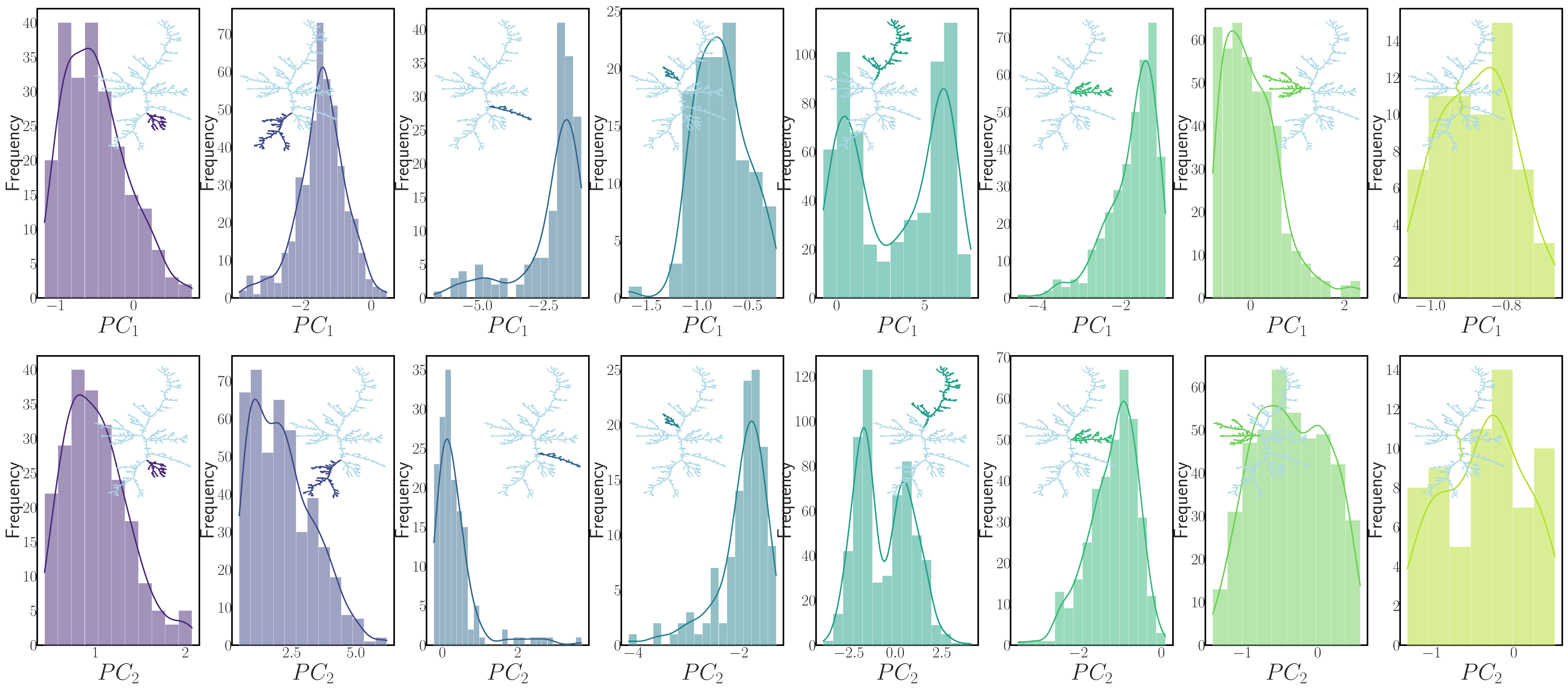}
  \centering
\caption{The distributions of the 2 principal components according to the significant branches seen in Fig. \ref{fig:MST_Branches_V167}, as noted in the inset. All panels are plotted in the corresponding scale for $PC_1$ and $PC_2$. The last panel in each row shows the distributions for the main trunk.}
       \label{fig:dist-var1}
\end{figure*}

\begin{figure*}
  \includegraphics[width=0.98\textwidth]{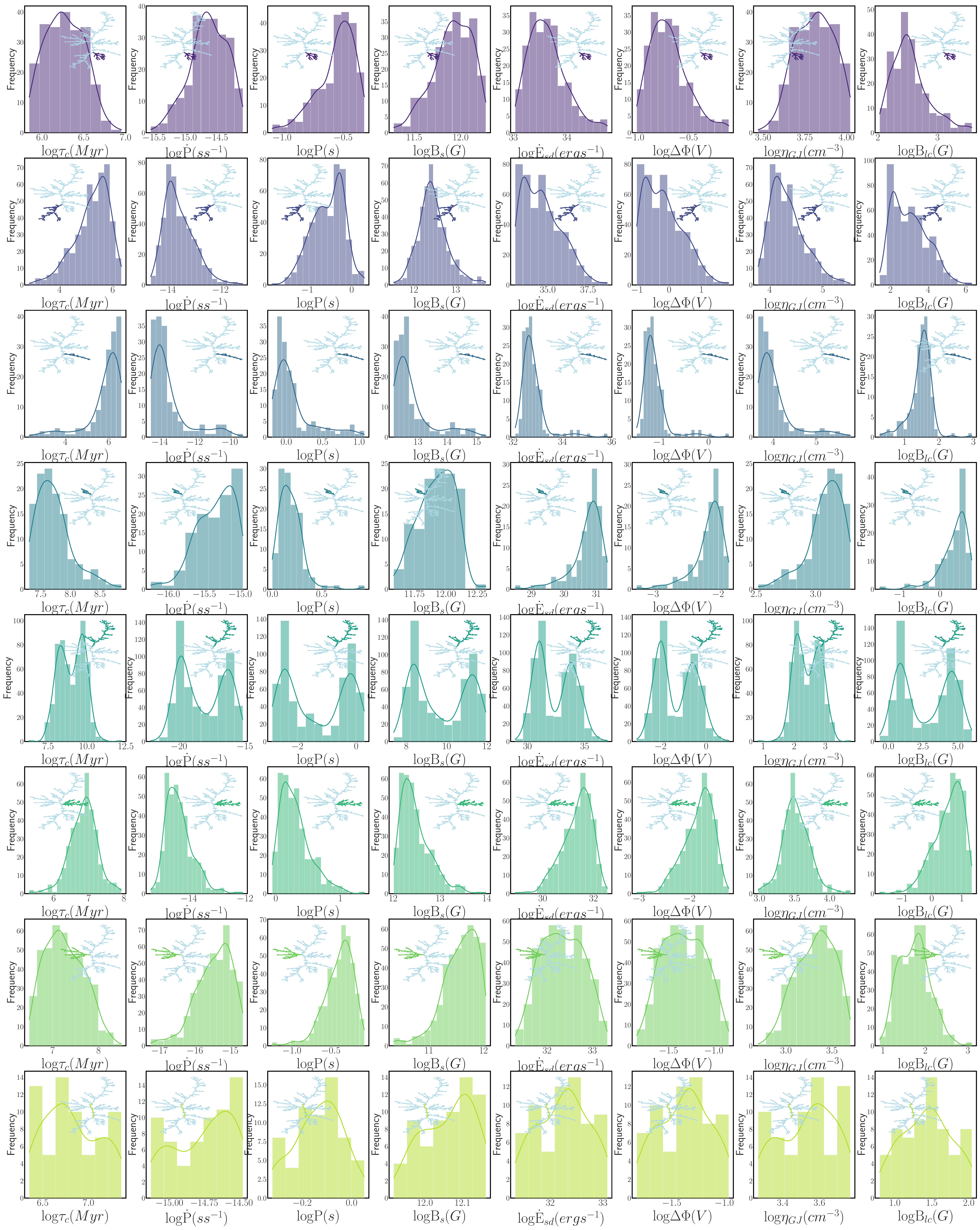}
  \centering
\caption{The distributions of the logarithm of the 8 intrinsic variables considered to build the principal components, arranged according to the significant branches (1 for each row) seen in Fig. \ref{fig:MST_Branches_V167}, as noted in the inset. The variables shown are: spin period, spin period derivative, characteristic age, spin-down energy loss rate, surface magnetic flux density, the magnetic field at the light cylinder surface electric voltage, and Goldreich-Julian current density. All panels are plotted in the corresponding scale for each of the magnitudes. The last row shows the distributions for the main trunk.}
       \label{fig:dist-var2}
\end{figure*}

Taking into the betweenness centrality distribution of Fig. \ref{Distributions_betweenness} and Eq.~(\ref{eq:outliers}), we obtain that 10\% of the nodes are outliers of the betweenness centrality distribution. These nodes fulfill the necessary conditions to be part of the main trunk. The use of the iterative method described provides a significance threshold of 3.8\%, i.e., each of the branches has to have at least 3.8\% of the total number of nodes, or at least a hundred pulsars. The significant branches so identified are shown in Fig. \ref{fig:MST_Branches_V167}.
The branches identified bear resemblance with the ones one would simply point by hand if looking at the MST. Some of these branches were already used as examples in Section 4 of our earlier work \citep{MST-1} when commenting on the pulsar tree as a descriptive tool.
The top and bottom branches of Fig. \ref{fig:MST_Branches_V167} roughly correspond to binary pulsars, and to the more energetic isolated pulsars of the sample (the youngest and those with the highest light cylinder magnetic field pulsars are located towards the end of this branch), respectively. Rightwards departing branches characterize by increasing values of surface magnetic fields, ending with magnetars at their extremes. The outgoing leftward branch contains the oldest isolated pulsars. Fig. \ref{fig:pt-examples} shows examples of the distribution of variables for some of the significant branches of the pulsar tree, as can be extracted from the online tool at {\url{http://www.pulsartree.ice.csic.es}}. It is clear that the significant branches separate different physical properties. To emphasize this point, Fig. \ref{fig:dist-var1} compares the principal components corresponding to each branch, as defined by \cite{MST-1}, showing differences in their distribution. We also note the case of the fifth panel from the left in Fig. \ref{fig:dist-var1}, showing the distribution of the branch containing the binary pulsars. This branch is large and mixes pulsars in binaries with isolated ones, and as such, it shows a double peak structure in the distribution, which happens in no other branch of the MST. Looking at the latter, the bottom part of the branch of this large structure (where we find a degree 4 node and two leftward departing branches)
is where the long-period, isolated pulsars are located. This can also be clearly seen using the Pulsar Tree Web. One can also gather that these two branches contain 118 pulsars in total (73, and 45 in each). Thus none of these branches is in excess of the significance threshold, and according to the definition, they cannot be classified as individually significant as of yet. We find however that by doing the KS test to test the null hypothesis between the corresponding distribution of the whole structure, of the individual branches, and all other branches we determine the null hypothesis is rejected. This is then a case where we foresee that an increased number of pulsars will likely join these branches into one generating one additional significant branch of the MST, or increase each of the branches' number of nodes to make them individually significant. Fig. \ref{fig:dist-var2} shows the distributions of each of the underlying physical magnitudes considered to construct the principal components. In this latter case, some of the individual variables seem to have similar distributions in the different branches. However, according to the KS test, the branches (that were chosen from $PC_1$) also show rejection of the null hypothesis when considering the individual magnitudes directly. All the information separately hosted in these variables is captured by $PC_1$ itself. The use of $PC_1$, therefore, simplifies and accelerates the comparison.
This seems to be a general result along the MSTs we have analyzed, although we cannot rule out a priori that some branches could be similar under the KS test of a particular intrinsic magnitude despite being dissimilar (always in terms of the null hypothesis) under the KS test of the principal components. We note the bimodality appearing in one of the top branches of the MST. This happens because most of the binary pulsars are located in that part (see right panel of Fig. \ref{fig:pt-examples}) and therefore, the analyzed variables and thus the PCs are affected by the behavior of these binary systems.

\subsubsection{Simulations}

To give further credit to the branch separation achieved we have carried out simulations. Particularly, we have considered a random separation of the pulsar population into groups corresponding to the number of branches we identify, each with the respective number of pulsars. We thus created fake, random branches, not motivated by the MST. We have done this exercise more than 10$^5$ times, and in each instance, we computed the KS test among the random 'branches' so created. Statistical separating fake branches prove very difficult: Not in a single case of our simulations, we find that all 7 branches can be separated. In fact, in 55\% of the simulations, the test is not able to reject the null hypothesis for not even 2 random branches. 

\section{Evolution of the MST with the pulsar population }
\label{versions}

\begin{figure*}
  \includegraphics[width=1\textwidth]{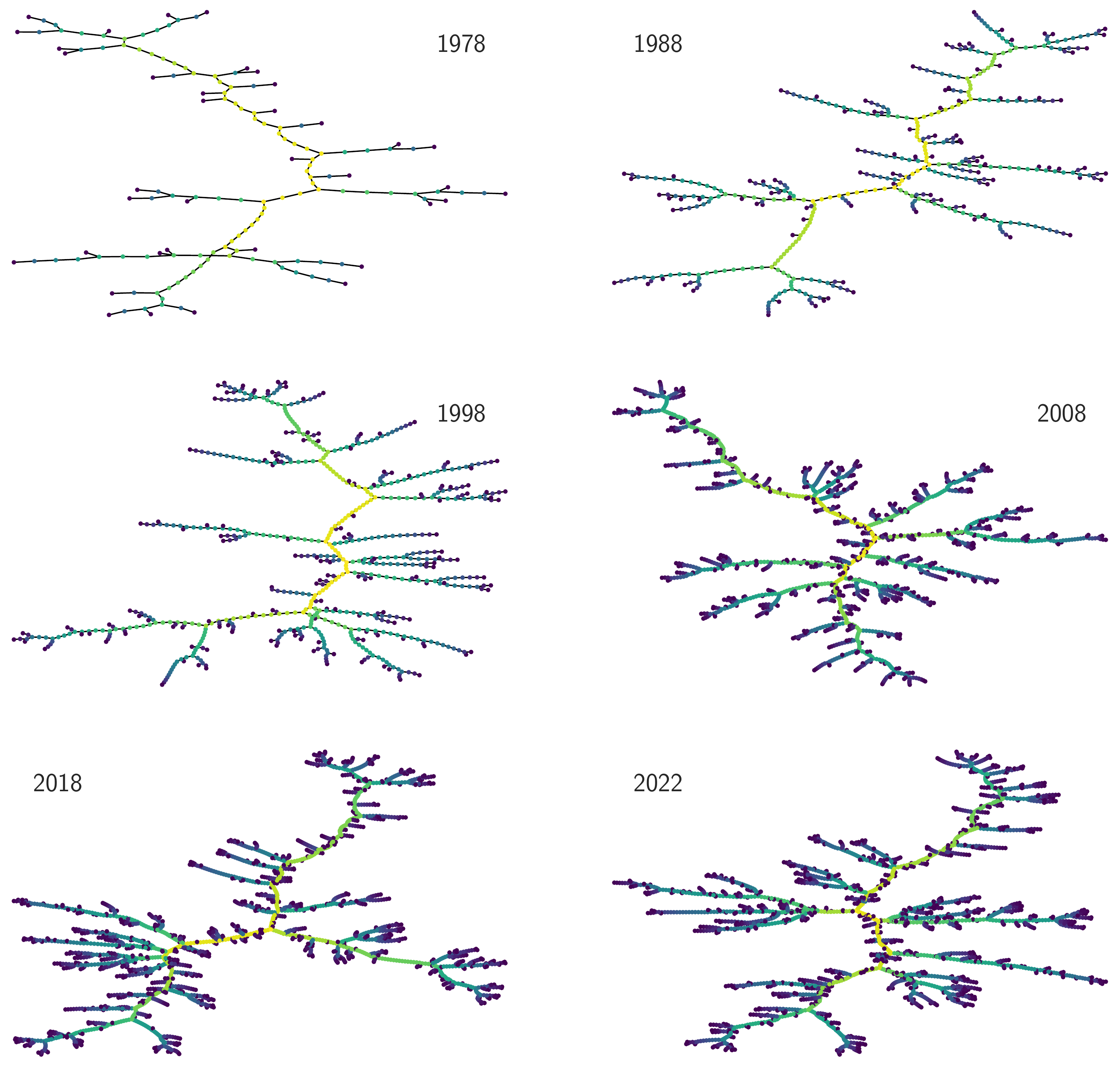}
  \centering
\caption{Representation of the $C_B$ values found after the application of Eq. (\ref{eq:betweenness_centrality}) in the MST graphs of the pulsar tree population along history, with nodes representing all pulsars known up to the different years according to the legend (up to December 31st of the prior year, and up to the date of version 1.67 of the ATNF catalog in March 2022). Similar to Fig. \ref{fig:BetCen_ModelVersion}, and in the same scale, the color map represents the highest $C_B$ values in yellow colors and increases the darkness of the paint to dark purple for those values that reach zero.
}
       \label{Betweenness_MST_versions}
\end{figure*}

\begin{figure*}
  \includegraphics[width=1\textwidth]{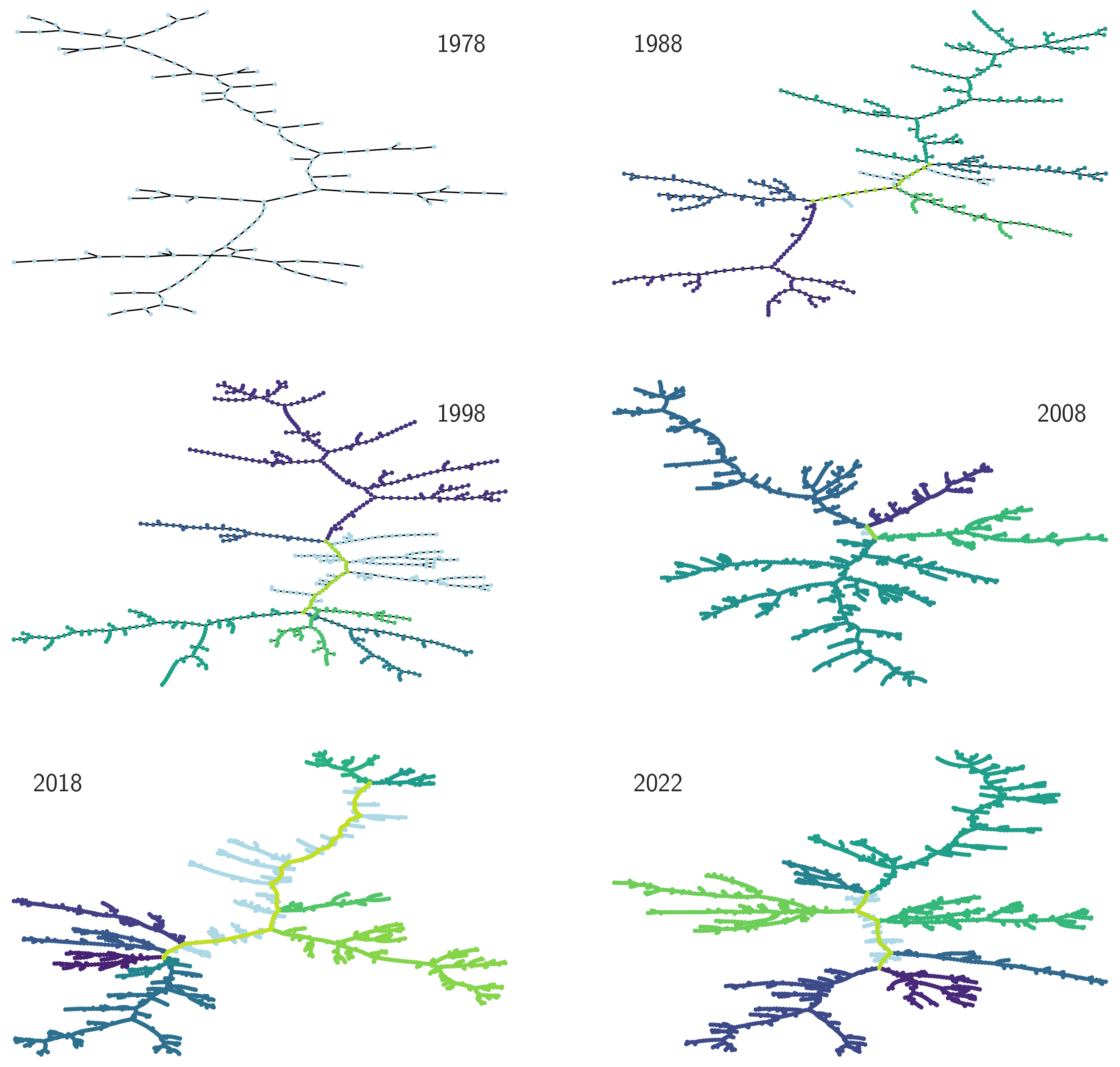}
  \centering
\caption{Significant branches and trunks (see \S \ref{versions}) for each of the MSTs corresponding to the time indicated in the legend. No part of the MST in 1978 is indicated because the application of Eq. (\ref{eq:outliers}) does not find nodes that can represent the trunk (there are no outliers of betweenness centrality). The different branches for each epoch are assigned a random color (MSTs are not color-coded for intercomparison). The trunk is always shown in each case with the most intense green color. The number of branches from 1988 to 2022 is 5, 5, 4, 9, and 7 respectively.
}
       \label{fig:Branches_trunk_versions}
\end{figure*}

We now turn to consider how our methods apply to the evolution of the pulsar population throughout history. This analysis is illustrative of its application, but we remark that as the pulsar population builds up in history with the number of pulsars known, thus a larger sample must necessarily be more significant for population analysis. We shall consider the pulsars known until the years 1978, 1988, 1998, 2008, and 2018, and the current population as described above, 2022. These sets contain 147, 439, 662, 1660, 2267, and 2509 pulsars, respectively. Application of Eq. (\ref{eq:outliers}) establishes a maximum percentage of values that can be considered outliers for each of these samples, resulting in  0\%, 7\%, 9\%, 9\%, 13\%, and 10\%, respectively. We recall that the nodes conforming to the trunk in each of the catalogs must have a $C_B$ value that makes them outliers of the corresponding betweenness centrality distribution. Fig. \ref{Betweenness_MST_versions} shows the pulsar tree along history, marking the betweenness centrality values in the same scale of Fig.~\ref{fig:BetCen_ModelVersion}. Note that due to the topological conditions imposed for the definition of the main trunk (see \S\ref{conditions}; i.e., that the trunk is a path, and especially, that the trunk must start and end in nodes of degree 3 or more that give rise to branches with at least a given a number of nodes) the size of the branches that can be a priori reached also has a maximum value beyond which these trunk conditions cannot be fulfilled. This maximum value is 9.7\%, 9.3\%, 8.9\%, 7\%, and 14.5\% respectively from 1988 onwards; the significance threshold obtained is consistently lower than them in all epochs.
Interestingly, due to the small size of the sample, the MST of the population of pulsars known until 1978 does not provide any outlier in the distribution of betweenness centrality. Correspondingly, that MST is less structured. The significance thresholds are 0\%, 6.2\%, 5.8\%, 8.6\%, 2.8\%, and 3.8\%, respectively. The number of significant branches along the history is 0, 5, 5, 4, 9, and 7, respectively. It is interesting to see that the number of branches is similar throughout history, despite the introduction of up to 20 times more pulsars than those known up to 1978. These are shown in Fig. \ref{fig:Branches_trunk_versions}.
We emphasize, as stated before, that these MSTs are still frames of the knowledge gathered from the pulsar population known up to each year quoted, where even whole classes of pulsars were undiscovered. To proceed further we shall focus only on the most complete population, based on the current catalog.

\section{Pulsar classes appearance: togetherness and growth of the current significant branches}
\label{history}

\begin{figure*}
  \includegraphics[width=1\textwidth]{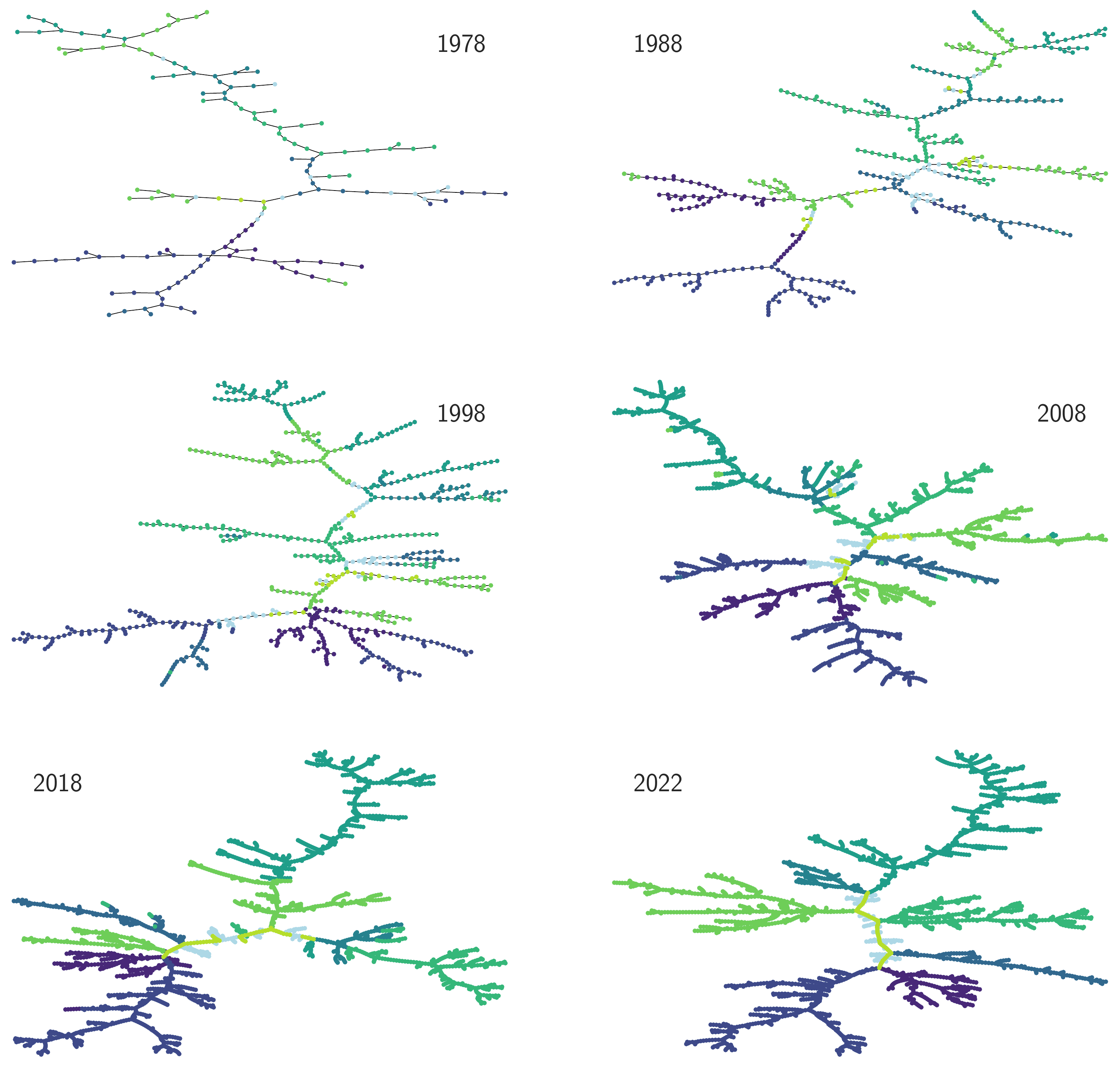}
  \centering
\caption{Projection of the 2022 branches into the MSTs corresponding to each epoch, following a chronological order as indicated within each panel. The last panel corresponds to that shown in Fig. \ref{fig:MST_Branches_V167}. 
}
       \label{fig:Branches_trunk_v167}
\end{figure*}

\begin{figure*}
  \includegraphics[width=0.9\textwidth]{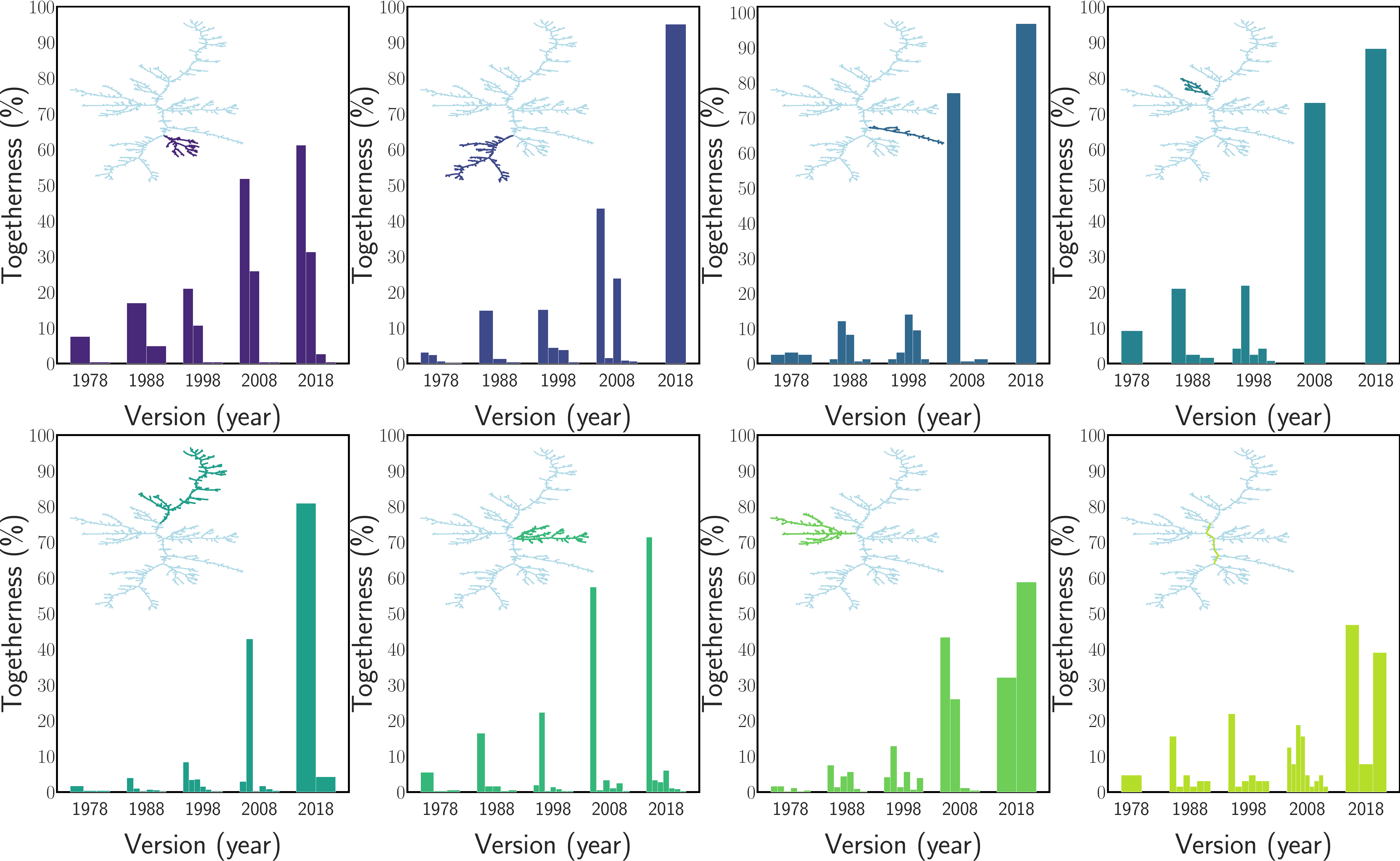}
  \centering
  \caption{Percentage of nodes with respect to the number of nodes of each significant branch today, according to the marking in the inner MST depicted in each panel that was already together in an MST at earlier times. When there were different groups of nodes in the earlier versions of the MST (see Fig. \ref{fig:Branches_trunk_v167}) corresponding to the same current branch, different bars are shown; whereas, when the nodes in a significant branch today were located all together in a single group, only a bar appears.
  }
  \label{fig:Branches_appearance}
\end{figure*}

\begin{figure*}
  \includegraphics[width=.9\textwidth]{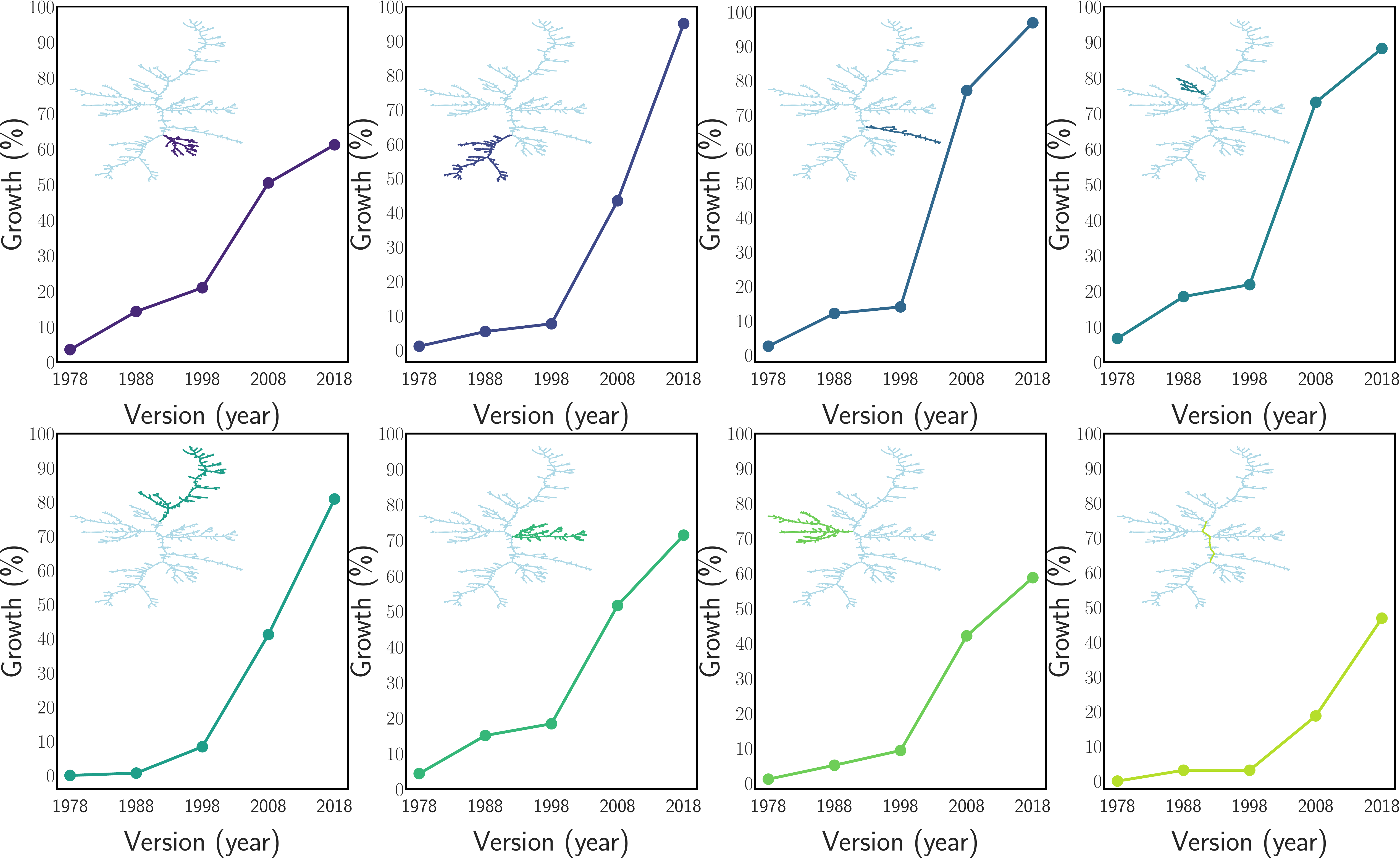}
  \centering
  \caption{Percentage of nodes with respect to the number of nodes of each significant branch today, according to the marking in the inner MST depicted in each panel that stays together throughout history, after appearing. To obtain these percentages we have taken into account the best representation (the largest bar for each branch in 2018 as seen in Fig. \ref{fig:Branches_appearance} in each panel) and analyzed how this group has grown from the previous period, repeating the process until 1978.
}
  \label{fig:percentage_of_growth}
\end{figure*}

Fig. \ref{fig:Branches_trunk_v167} shows how the nodes in each of the current branches were distributed in the earlier MSTs since their appearance. Current branches get cut or mix into 2 or more pieces and were more distributed in the MST the smaller the total number of pulsars is. As time goes by, there is a process of convergence for all the current branches, which is in most cases even visually obvious because most of the nodes in the current branches are connected to one another earlier on as well, i.e. when already discovered, they are adjacent nodes also in earlier versions of the catalog. To quantify this process, we measure the percentage of {\it togetherness}: how many of the current nodes in a given branch (or the trunk) were together at earlier times. This is shown in Fig. \ref{fig:Branches_appearance}. For those cases in which a given branch cuts into smaller pieces (e.g., its nodes populate two branches of the previous incarnation of the catalog) we shall also consider the group containing the largest number of nodes, as this is necessarily more representative of the final branch. Starting from a given branch in the current catalog (2022), we shall count how many of the nodes were existing and were located together in a branch of the previous version (2018). Starting from the latter set, we checked how many of those were existing and were already together in the catalog before it (1998), and so on and so forth. Differently to the measure of togetherness, {\it growth} always concerns the same pulsars, those that, when appearing, stay together onwards in time in all subsequent MSTs, being joined by others until the size of the current branch is achieved. This discounts the process of convergence of different groups into the same current branch, as shown in Fig. \ref{fig:Branches_appearance}. Fig. \ref{fig:percentage_of_growth} shows these results. Thus, we see that all branches in the current MST not only join groups of pulsars that were earlier separated (Fig. \ref{fig:Branches_appearance}) but also have a core group of pulsars that stay together along all catalogs (Fig. \ref{fig:percentage_of_growth}).

\section{Conclusions}
\label{conclu}
A new class of pulsars is, conceptually, a new set of pulsars whose properties are distinct from the others. In this work, we have introduced a methodology based on the graph theory-motivated pulsar tree and the underlying principal components analysis of the pulsar population \citep{MST-1} that allows studying this appearance. In general terms (since it may find applications for other problems beyond pulsars) we have introduced a methodology to use the MST to qualify the nodes forming it. The logic is as follows:
\begin{enumerate}
\item Produce the MST based on the Euclidean distance of the variables containing the full variance of the population as determined from principal components analysis.
\item Compute betweenness centrality for all nodes, produce its distribution, and find out which nodes are outliers. 
\item Compute the significance threshold, defined as the minimum percentage of the total population of pulsars for which the branches are statistically distinct (measured by a KS test), maximizing the length of the trunk.
\item If having former incarnations of the sample, with a reduced number of nodes (in our cases, those provided by the historical discovery of pulsars), compute togetherness and growth to see convergence stability and establishment of significant branches.
\end{enumerate}
The former methodology allows us to quantitatively define the trunk and the significant branches of any tree, for which the nodes
have statistically different distributions of their principal components. In the pulsar case, using the principal components associated with the intrinsic variables, the significant branches can be directly associated with `classes', or at least, to connections among the respective nodes, which quantitatively separate them from others in different locations of the tree. The study of the population evolution along history under this methodology allows us to see how our knowledge gets completed, and what initially were apparently different pulsars related to each other. New sets of data, e.g., a future incarnation of the catalog bringing a significant number of new pulsars will also make the current MST change. We may see new branches appearing, with no or minimal projection onto the current 2022 catalog of pulsars, showing a brand new class. We may also see current non-significant sub-branches develop into full-fledged significant ones. The MST will provide a quantitative perspective on whether we are seeing something `new' or just `more of the same' when new pulsars are discovered, and, similarly, whether currently unknown pulsars will connect groups that are currently dislocated. 
A spinoff application of our method could also be the testing of population synthesis models. If we have the output of a pulsar population synthesis model giving us the same number of pulsars as in the real catalog, we can build the MST  of this synthesized population, analyze the distribution of the degrees of the nodes, and use our method as in the real case to determine the MST and significant branches. A comparison of the distribution of the degrees of the nodes may directly assess whether we are looking at similar MSTs, via graph properties. Non-parametric tests of the distribution of the principal components of the simulated branches 
compared with the real ones will be a direct indication of the goodness of the underlying population synthesis model. We hope to explore this in detail in future work.

\section*{Acknowledgements}
This work has been supported by the grant PID2021-124581OB-I00 funded by MCIN/AEI/10.13039/501100011033. C.R. is funded by the Ph.D. FPI fellowship PRE2019-090828 acknowledges the graduate program of the Universitat Aut\`onoma of Barcelona. This work was also supported by the Spanish program Unidad de Excelencia María de Maeztu CEX2020-001058-M.

\section*{Data availability}

The data underlying this article are available in "The pulsar tree" web at the Institute of Space Sciences (ICE, CSIC) \url{http://pulsartree.ice.csic.es}.

\bibliographystyle{mnras}
\bibliography{biblio}

\begin{thebibliography}{}
\makeatletter
\relax
\def\mn@urlcharsother{\let\do\@makeother \do\$\do\&\do\#\do\^\do\_\do\%\do\~}
\def\mn@doi{\begingroup\mn@urlcharsother \@ifnextchar [ {\mn@doi@}
  {\mn@doi@[]}}
\def\mn@doi@[#1]#2{\def\@tempa{#1}\ifx\@tempa\@empty \href
  {http://dx.doi.org/#2} {doi:#2}\else \href {http://dx.doi.org/#2} {#1}\fi
  \endgroup}
\def\mn@eprint#1#2{\mn@eprint@#1:#2::\@nil}
\def\mn@eprint@arXiv#1{\href {http://arxiv.org/abs/#1} {{\tt arXiv:#1}}}
\def\mn@eprint@dblp#1{\href {http://dblp.uni-trier.de/rec/bibtex/#1.xml}
  {dblp:#1}}
\def\mn@eprint@#1:#2:#3:#4\@nil{\def\@tempa {#1}\def\@tempb {#2}\def\@tempc
  {#3}\ifx \@tempc \@empty \let \@tempc \@tempb \let \@tempb \@tempa \fi \ifx
  \@tempb \@empty \def\@tempb {arXiv}\fi \@ifundefined
  {mn@eprint@\@tempb}{\@tempb:\@tempc}{\expandafter \expandafter \csname
  mn@eprint@\@tempb\endcsname \expandafter{\@tempc}}}

\bibitem[\protect\citeauthoryear{Brandes}{Brandes}{2001}]{Brandes}
Brandes U.,  2001, \mn@doi [The Journal of Mathematical Sociology]
  {10.1080/0022250X.2001.9990249}, 25, 163

\bibitem[\protect\citeauthoryear{Freeman}{Freeman}{1977}]{original}
Freeman L.~C.,  1977, Sociometry, 40, 35

\bibitem[\protect\citeauthoryear{{Garc{\'\i}a}, {Torres}  \&
  {Patruno}}{{Garc{\'\i}a} et~al.}{2022}]{MST-1}
{Garc{\'\i}a} C.~R.,  {Torres} D.~F.,   {Patruno} A.,  2022, \mn@doi [\mnras]
  {10.1093/mnras/stac1997}, \href
  {https://ui.adsabs.harvard.edu/abs/2022MNRAS.515.3883G} {515, 3883}

\bibitem[\protect\citeauthoryear{{Gower} \& {Ross}}{{Gower} \&
  {Ross}}{1969}]{Gower1969}
{Gower} J.~C.,  {Ross} G. J.~S.,  1969, Journal of the Royal Statistical
  Society. Series C (Applied Statistics), 18, 54

\bibitem[\protect\citeauthoryear{{H.~E.~S.~S. Collaboration}
  et~al.,}{{H.~E.~S.~S. Collaboration} et~al.}{2018}]{HESSPWNe}
{H.~E.~S.~S. Collaboration} et~al., 2018, \mn@doi [\aap]
  {10.1051/0004-6361/201629377}, \href
  {https://ui.adsabs.harvard.edu/abs/2018A&A...612A...2H} {612, A2}

\bibitem[\protect\citeauthoryear{Jones}{Jones}{1994}]{Ftests}
Jones D.~H.,  1994, Journal of Educational and Behavioral Statistics, 19, 304

\bibitem[\protect\citeauthoryear{Kruskal}{Kruskal}{1956}]{Kruskal1956}
Kruskal J.~B.,  1956, \mn@doi [Proc. Amer. Math. Soc.]
  {10.1090/S0002-9939-1956-0078686-7}, 7, 48

\bibitem[\protect\citeauthoryear{Lehmann}{Lehmann}{2012}]{KS-test5}
Lehmann E.~L.,  2012, in , Selected Works of EL Lehmann.
Springer, pp 373--390

\bibitem[\protect\citeauthoryear{{Manchester}, {Hobbs}, {Teoh}  \&
  {Hobbs}}{{Manchester} et~al.}{2005}]{ATNF-Catalog}
{Manchester} R.~N.,  {Hobbs} G.~B.,  {Teoh} A.,   {Hobbs} M.,  2005, \mn@doi
  [\aj] {10.1086/428488}, \href
  {https://ui.adsabs.harvard.edu/abs/2005AJ....129.1993M} {129, 1993}

\bibitem[\protect\citeauthoryear{Moxley \& Moxley}{Moxley \&
  Moxley}{1974}]{Moxley1974}
Moxley R.~L.,  Moxley N.~F.,  1974, Sociometry, 37, 122

\bibitem[\protect\citeauthoryear{{Pearson}}{{Pearson}}{1901}]{Pearson1901}
{Pearson} K.,  1901, \mn@doi [The London, Edinburgh, and Dublin Philosophical
  Magazine and Journal of Science] {10.1080/14786440109462720}, 2, 559

\bibitem[\protect\citeauthoryear{Scholz \& Stephens}{Scholz \&
  Stephens}{1987}]{anderson}
Scholz F.~W.,  Stephens M.~A.,  1987, Journal of the American Statistical
  Association, 82, 918

\bibitem[\protect\citeauthoryear{{Shlens}}{{Shlens}}{2014}]{Shlens2014}
{Shlens} J.,  2014, arXiv e-prints, \href
  {https://ui.adsabs.harvard.edu/abs/2014arXiv1404.1100S} {p. arXiv:1404.1100}

\bibitem[\protect\citeauthoryear{{Tarjan}}{{Tarjan}}{1983}]{Tarjan1983}
{Tarjan} R.~E.,  1983, Data structures and network algorithms.
Society for Industrial and Applied Mathematics

\bibitem[\protect\citeauthoryear{Tukey}{Tukey}{1977}]{Tukey}
Tukey J.~W.,  1977, Exploratory Data Analysis.
Addison-Wesley, Reading, Massachusetts

\bibitem[\protect\citeauthoryear{{Wilson}}{{Wilson}}{2010}]{Wilson2010}
{Wilson} R.~J.,  2010, Introduction to graph theory.
Pearson

\bibitem[\protect\citeauthoryear{Wolfe}{Wolfe}{2012}]{KS-test4}
Wolfe D.~A.,  2012, in Rojo J.,  ed., , Selected Works of E. L. Lehmann.
Springer US, Boston, MA, pp 1101--1110, \mn@doi{10.1007/978-1-4614-1412-4_96},
  \url {https://doi.org/10.1007/978-1-4614-1412-4_96}

\bibitem[\protect\citeauthoryear{Yadolah}{Yadolah}{2008}]{KS-test3}
Yadolah D.,  2008, in , The Concise Encyclopedia of Statistics.
Springer New York, New York, NY, pp 283--287,
  \mn@doi{10.1007/978-0-387-32833-1_214}, \url
  {https://doi.org/10.1007/978-0-387-32833-1_214}

\makeatother
\end{thebibliography}

\section*{Appendix
\label{sec:appendix}}

\begin{figure*}
  \includegraphics[width=1\textwidth]{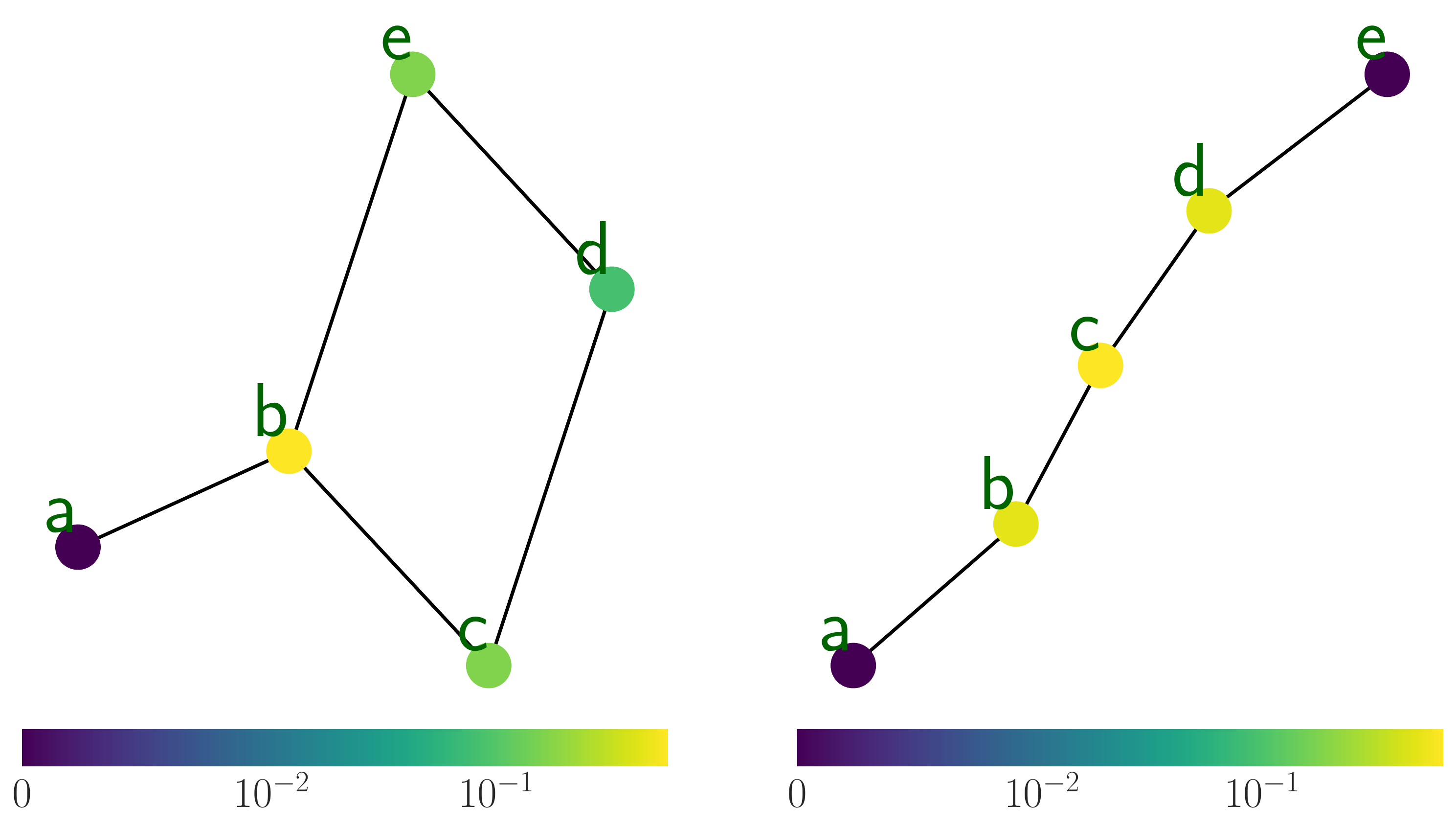}
  \centering
  \caption{
  Left: A graph $G(5,5)$ with a cycle composed of nodes $(b, c, d, e)$. The node with the largest $C_B$ is $v=b$ whose value would be 0.58, while $v=a$ would take the value of 0. Right: A graph $T(5,4)$ is the MST coming from $G$. The node with the largest $C_B$ is $v=c$ whose value is 0.66, while $v=a$ and  $v=e$ would take the value of 0. Similar to Fig. \ref{fig:BetCen_ModelVersion}, and in the same scale, the color map represents the highest $C_B$ values in yellow colors with increasing  darkness for  values toward zero.
}
  \label{CB_appendix}
\end{figure*}

We provide here a simple example of betweenness centrality computation. To calculate the coefficient $C_b$ by applying Eq. (\ref{eq:betweenness_centrality}) on a graph, we recall to take into account the following considerations:
 \begin{itemize}

    \item Select all pairs of nodes ${s,t}$. One can already exclude neighboring (adjacent) ones, as there cannot be another node in between them.
    \item Select all paths within the graph joining the selected nodes, eliminating duplicities (i.e., $\sigma_{s,t} = \sigma_{t,s}$). 
    \item Each node, $v$, appearing on these paths will be assigned the numerical value 1, i.e., $\sigma_{s,t}(v)$=1.
    \item Consider the shortest paths (geodesic distance) between the nodes $s$ and $t$ under consideration. If the geodesic distance between the nodes in question could be reached by traversing different paths, due to the existence of cycles, $\sigma_{s,t}$ would take a value equal to the number of such paths. 
    \item Make the total sum of the values according to the previous steps. We will normalize the above value by multiplying the result obtained by $2/(N-1)(N-2)$.
 \end{itemize}
We apply these steps to two different examples, see Fig. \ref{CB_appendix}, to see how the arrangement of nodes in the graph affects $C_B$. The left panel shows a graph $G(5,5)$ whose highest coefficient node is the one denoted as $b$. The node $b$ appears in the shortest paths $\sigma_{a,c}(b)$, $\sigma_{a,d}(b)$, $\sigma_{a,e}(b)$ and $\sigma_{c,e}(b)$ so we assign the value 1 for each of the times $b$ appears in the shortest path(s) between these pairs. Of those pairs, on the other hand, $\sigma_{a,d}$ and $\sigma_{c,e}$ have a geodesic distance that reaches the same value by two different paths, therefore, a value of 2 will be assigned to each of them. Concurrently, $\sigma_{a,c}$ and $\sigma_{a,e}$ will be equal to 1 because there is only one shortest path to link the corresponding pairs. So the sum total for node $b$ is $\mathrm{1+\frac{1}{2}+\frac{1}{2}+\frac{1}{2}+1}$=3.5 and if we normalize ($N=5$) the final value is $C_B(b)$=0.58. The second example (right panel of Fig. \ref{CB_appendix}) represents the MST $T(5,4)$ of the graph $G$. As an MST does not contain loops, the geodesic distance between a pair of nodes can only be reached through a single path. This path will therefore always be the shortest one so that  $\sigma_{s,t}$ will always be 1. Then, $\sigma_{s,t}(v)$ will at most be equal to 1 for each node $v$. Focusing again on node $b$, it appears in the following paths $\sigma_{a,c}(b)$, $\sigma_{a,d}(b)$ and $\sigma_{a,e}(b)$, which will all contribute 1 to the sum. The final sum is then 3, which normalized as defined above is 0.5.

\label{lastpage}
\end{document}